\documentclass[aps,pre,twocolumn,nofootinbib,floatfix,showpacs]{revtex4}  

\usepackage{graphicx} 
\newcommand {\be}{\begin{equation}}
\newcommand {\ee}{\end{equation}}
\newcommand {\ba}{\begin{eqnarray}}
\newcommand {\ea}{\end{eqnarray}}

\begin{document}

\title{Macroscopic evidence of microscopic dynamics in the Fermi-Pasta-Ulam oscillator chain from nonlinear time series analysis}

\author{M. Romero-Bastida}
\email{rbm@xanum.uam.mx}
\author{D. Casta\~neda}
\author{E. Braun}
\affiliation{Departamento de F\'\i sica, Universidad Aut\'onoma Metropolitana Iztapalapa, Apartado Postal 55--534, Distrito Federal 09340, M\'exico} 

\date{\today}

\begin{abstract}
The problem of detecting specific features of microscopic dynamics in the macroscopic behavior of a many-degrees-of-freedom system is investigated by analyzing the position and momentum time series of a heavy impurity embedded in a chain of nearest-neighbor anharmonic Fermi-Pasta-Ulam oscillators. Results obtained in a previous work~[M. Romero-Bastida, Phys. Rev. E {\bf69}, 056204 (2004)] suggest that the impurity does not contribute significantly to the dynamics of the chain and can be considered as a probe for the dynamics of the system to which the impurity is coupled. The ($r,\tau$) entropy, which measures the amount of information generated by unit time at different scales $\tau$ of time and $r$ of the observable, is numerically computed by methods of nonlinear time-series analysis using the position and momentum signals of the heavy impurity for various values of the energy density $\epsilon$ (energy per degree of freedom) of the system and some values of the impurity mass $M$. Results obtained from these two time series are compared and discussed.
\end{abstract}


\maketitle

\section{Introduction}

The physically relevant question of how the underlying microscopic dynamics of many-particle systems is related to the observed macroscopic behavior is still not satisfactorily answered. Since the main features of the latter are accounted for by the microscopic dynamics, which is chaotic for most systems of interest, it is natural to assume that some characteristics of the chaotic microscopic dynamics should be detected, in principle, at the macroscopic level of description. Only then it would be possible to assess the relevance of the microscopic dynamics on macroscopic behavior.

Empirical evidence suggesting the existence of microscopic chaos on a molecular scale has been presented in an experiment on the position of a Brownian particle (BP) in a fluid~\cite{Gaspard}. The measurements were made at regular time intervals and the experimental time series data was then interpreted using standard techniques of time series analysis, suggesting a positive lower bound on the Kolmogorov-Sinai entropy per unit time $h_{_{KS}}$, hence, microscopic chaos. However, a similar bound has been obtained with computer experiments on the nonchaotic Eherenfest wind-tree model where a single particle diffuses in a plane due to collisions with randomly placed, fixed, oriented square scatterers~\cite{Dettmann1}. Later on it was found that the chaotic Lorentz model, which has circular scatterers, has a diffusive behavior that is indistinguishable from the one exhibited by the Eherenfest model~\cite{Dettmann2}. Furthermore, a class of one-dimensional maps has also been reported which present normal diffusivelike behavior in the absence of chaos, just as in the Eherenfest model~\cite{Cecconi}. These results have rendered doubts about the conclusion that microscopic chaos has been experimentally detected. The positive bound on $h_{_{KS}}$ of Ref.~\cite{Gaspard} has been attributed to the finiteness in both spatial and temporal resolution and the limited amount of data points~\cite{Dettmann1,Cencini}.

The results in all previous works have been obtained with only one of the BP dynamical variables, namely its position. However, there is a time scale in Brownian motion, not mentioned in the previous works, that is characterized by the relaxation time of the momentum autocorrelation function (MACF) of the BP. This function can be computed from the BP momentum time series. Significant variations in the BP momentum ocurr in this time scale, which is much smaller than the corresponding position time scale. Thus the BP momentum could, in principle, be used to probe the microscopic dynamics of the system. But the type of the models used in Refs.~\cite{Dettmann1,Dettmann2,Cecconi,Cencini} precludes a study in the momentum space from the very beginning, and therefore a different system is needed to address the problem at hand.

The Fermi-Pasta-Ulam (FPU) model, which is a one-dimensional chain of nearest-neighbor anharmonic oscillators, is a system that has been extensively studied over the past decades to test the way in which microscopic dynamics determines the macroscopic behavior. Now, if the FPU chain is coupled to a heavy impurity, the effects of the microscopic dynamics on the statistical behavior of the latter can be readily studied for a number of reasons. First, the position and momenta of all the oscillators and the heavy impurity are well defined variables. At variance with the Eherenfest and Lorentz models of Refs.~\cite{Dettmann1,Dettmann2}, in which the tracer particle interacts with the fixed scatterers through a discontinuous repulsive potential, the system composed of the FPU chain coupled to the heavy impurity is defined in terms of a more realistic continuous attractive potential. Furthermore, this is a many-particle system in which the statistical behavior of the heavy impurity is produced by the interaction with the oscillators of the chain, in sharp contrast with the models of Refs.~\cite{Cecconi,Cencini}, which are one-dimensional maps with only one dynamical variable, i.e. the position, defined in the model~\cite{Note1}.

It has been known for some time that, for the FPU chain, a single parameter, namely its energy per degree of freedom $\epsilon\equiv E/N$, controls its phase-space dynamics; that is, the system is weakly chaotic and thus has a dynamical behavior that, in the timescales so far studied, is almost identical to that of a chain of harmonic oscillators for low $\epsilon$ values, whereas it is strongly chaotic for high $\epsilon$ values~\cite{Pettini}. This behavior, for both low an high $\epsilon$ values, is unaltered by the inclusion of a heavy impurity~\cite{RomeroBraun}. The transition between these two dynamical regimes was first detected by means of a change in the scaling behavior of the largest Lyapunov exponent (LLE), which measures the exponential rate of divergence of two originally close trajectories in phase space, as $\epsilon$ is varied from a low to a high value for the FPU model~\cite{Pettini} and by a change in the scaling of the relaxation time of the MACF as a function of $\epsilon$ for the FPU chain coupled to a heavy impurity~\cite{RomeroBraun}.

In this work we will concentrate on the macroscopic manifestation of the transition from one dynamical regime to another, which can be readily studied in the macroscopic behavior of the heavy impurity embedded in a FPU chain. Our starting point comes from the observation that the Kolmogorov-Sinai entropy $h_{_{KS}}$, which has been computed from the complete microscopic dynamics of this system as a function of $\epsilon$, displays the aforementioned transition in the scaling behavior of $h_{_{KS}}$~\cite{Romero}. Since in the description of the microscopic dynamics the momentum of all the oscillators of the system is considered and the dynamics of the chain is not affected by the presence of the heavy impurity, we can ask whether some evidence of the above mentioned transition between dynamical regimes can be detected by applying the methods of nonlinear time series analysis to the momentum time series of the heavy impurity. The present work is a first step to answer the above posed question.

The plan of the paper is as follows. In Sec. II we briefly review the model and the methodology employed to obtain the position and momentum time series of the BP. In Sec. III we present the results for the position time series, which are consistent with those of Refs.~\cite{Dettmann1,Cencini}. In Sec. IV we apply the same methodology of the previous section to the momentum time series, with the result that there is a quantitative difference in the results obtained in the low and high $\epsilon$ regimes, and hence evidence of the microscopic dynamics of the oscillator chain. In Sec. V we discuss the results so far obtained. Conclusions are given in Sec. VI.

\section{The Model and its Numerical Investigation}

In terms of dimensionless variables, the Hamiltonian of the model considered in this work is
\be
H\!=\!\!\!\!\!\sum_{i=-N/2}^{N/2} \displaystyle\left[\frac{p_i^2}{2m_i}+\frac{1}{2}(x_{i+1}-x_i )^2 +\frac{1}{4}\beta (x_{i+1}-x_i )^4 \right]~\label{newham}
\ee
\noindent
with $m_i=1$ for $i\not=0$ and $m_0=M$; periodic boundary conditions are assumed ($x_{(N/2)+1}=x_{-N/2}$). The model, which will be named modified FPU (MFPU) model from now on, describes a system of one-dimensional $N$ coupled nonlinear oscillators of unit mass with nearest-neighbor interactions, displacements $\{x_i\}$, momenta $\{p_i\}$ and a central oscillator (impurity) of mass $M$ with displacement $x_0\equiv X$ and momentum $p_0\equiv P$. The value $\beta=0.1$ was used in the computation of all the numerical results hereafter reported. As initial conditions we choose the equilibrium value of the oscillators displacements, i.e. $x_i(0)=0$ for $i=-N/2,\ldots,N/2$. The momenta $\{p_i(0)\}$ were drawn from a Maxwell-Boltzmann distribution at a temperature consistent with a given value of the energy density $\epsilon$, which was chosen in the range $0.01\le\epsilon\le100$ with $M=40$, $60$, $80$, and $100$. Then the $2(N+1)$ equations of motion were numerically integrated to obtain the time evolution of the system, whose state is represented by the variable $\Gamma(t)=(\{x_i(t)\},\{p_i(t)\})\in\Re^{^{2(N+1)}}$. After thermal equilibrium between the impurity and the FPU chain with $N=300\,000$ unit mass oscillators is attained, the behavior of the heavy impurity is almost identical to a Brownian motion for all $\epsilon$ values studied~\cite{Note2}, as can be inferred from the values of the diffusion coefficient and the exponential fit to the MACF~\cite{RomeroBraun}. So, we can consider the heavy impurity as a genuine BP. It is in this regime that the position $\{X(t_{\alpha})\}$ and momentum $\{P(t_{\alpha})\}$ time series of the heavy impurity, with $t_{\alpha}\equiv\alpha\tau$ ($\alpha=1,\ldots,n$) and $\tau=1$ being the natural time unit (discretization step), were recorded for all $\epsilon$ and $M$ values considered over a time interval of $n=2\times10^5$ natural time units.

\section{Position time series analysis}

\begin{figure}
\includegraphics[width=0.98\linewidth,angle=0.0]{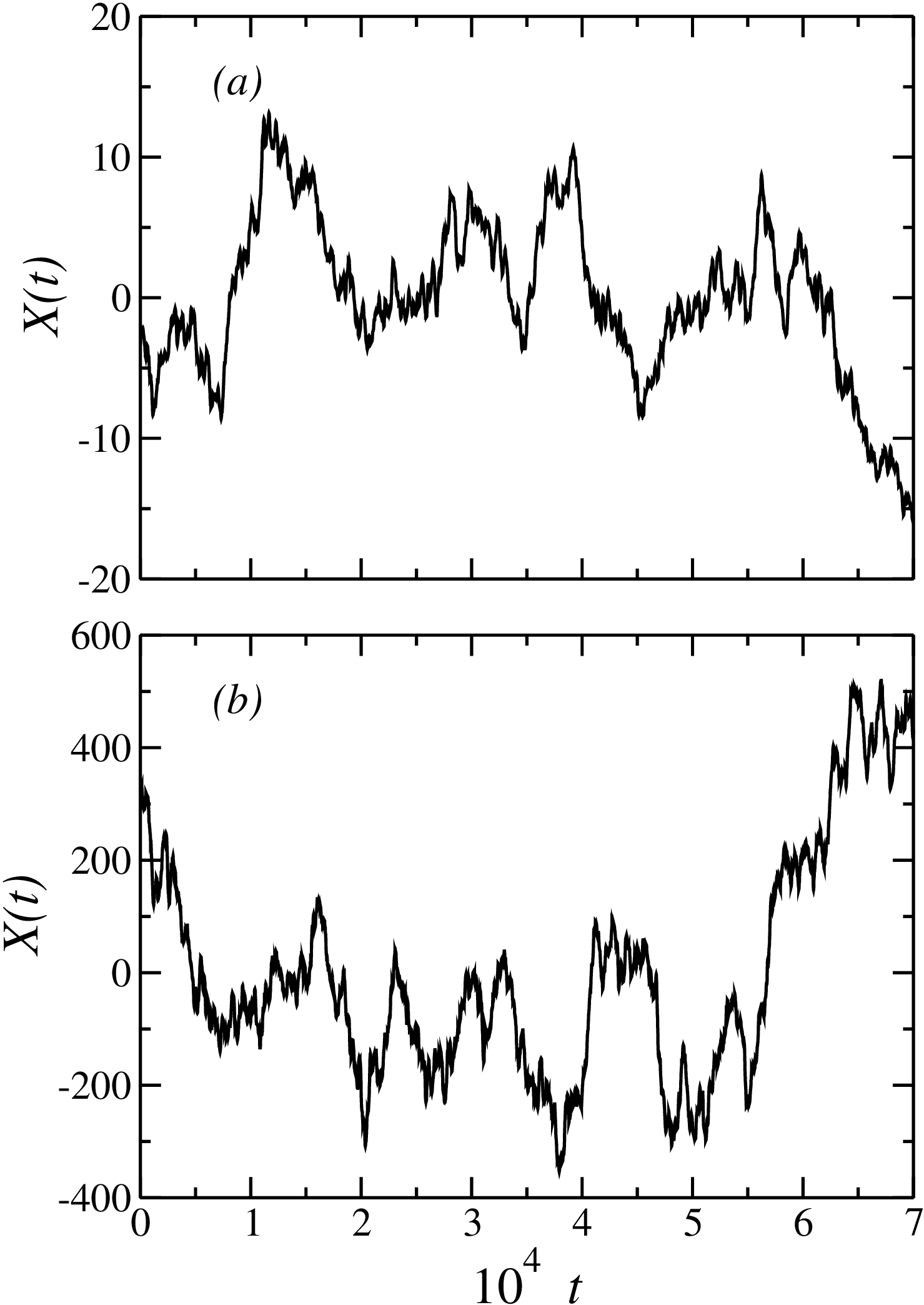}
\caption{(a) Position time record for $\epsilon=0.01$. The numerically computed diffusion coefficient is $D\approx0.0052$. The length of the complete data set is $2\times10^5$ and the data were sampled with $\tau=1$. (b) Position time record for $\epsilon=10$ with a numerically computed diffusion constant of $D\approx33.65$. Same total data set length and $\tau$ value as in (a). In both cases $M=100$.}
\label{fig:Pos}
\end{figure}

In Fig.\ \ref{fig:Pos} we show a segment of the complete position time series for $\epsilon=0.01$ and $10$ with $M=100$, which correspond to the weakly chaotic (i.e. almost periodic) and one instance of the strongly chaotic regime, respectively. We can readily appreciate that both signals are practically indistinguishable (except for the scale, which is controlled by the $\epsilon$ value), although they correspond to completely different dynamical regimes, as already mentioned. Furthermore, some quasirecurrences can be appreciated in the complete data series (not shown). This fact is a consequence of the finite $N$ value. Thus it can be inferred that the position of a BP is unsuitable as a probe to the microscopic dynamics of the chain.

\begin{figure}
\includegraphics[width=0.98\linewidth]{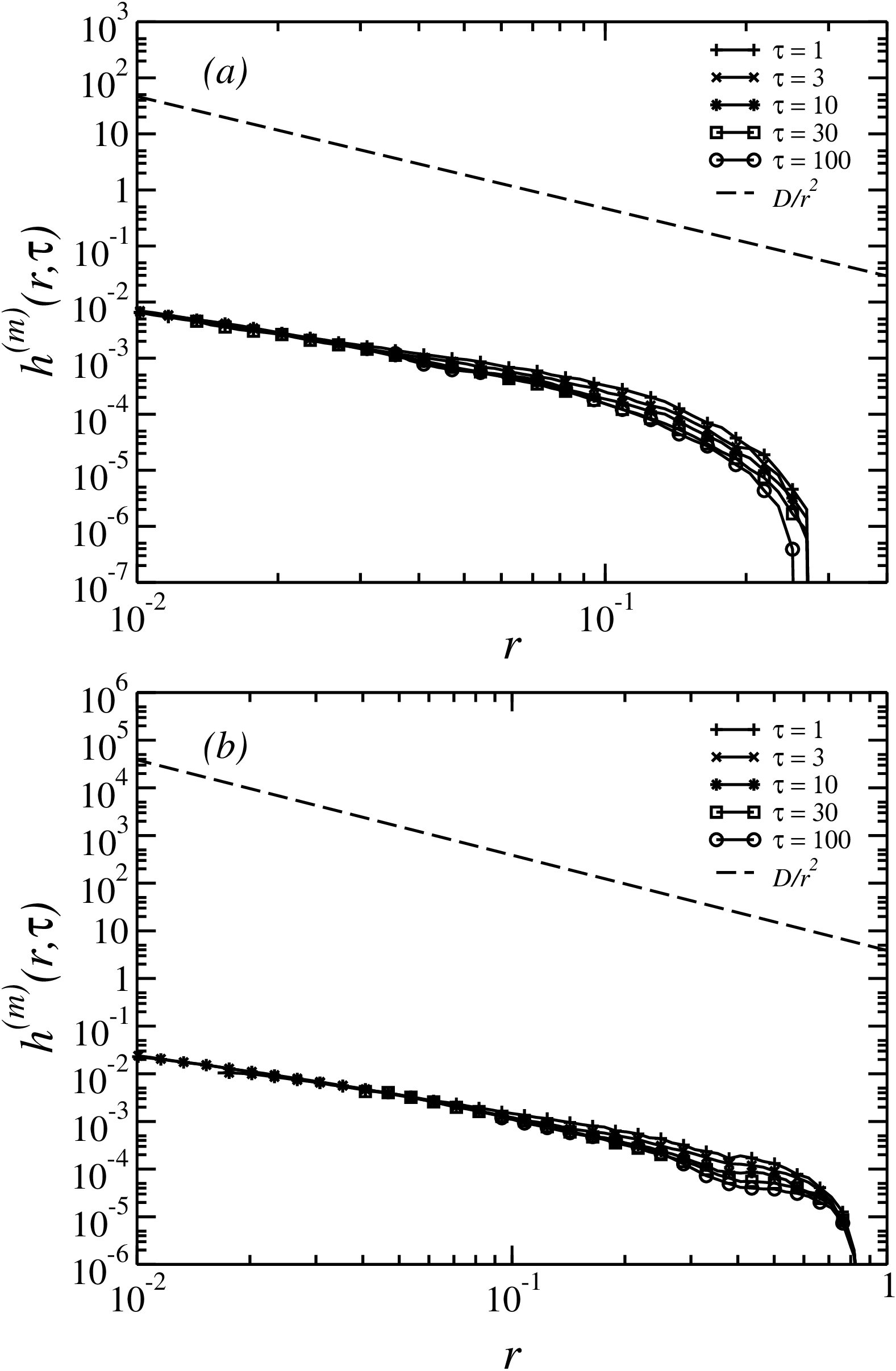}
\caption{$h^{(m)}(r,\tau)$ computed using $7\times10^4$ data points from the time series shown in Fig.\ \ref{fig:Pos}. The reported results are for an embedding dimension of $m=50$. (a) $\epsilon=0.01$ and (b) $\epsilon=10$. The two straight lines show the corresponding $D/r^2$ behavior.}
\label{fig:Hrtau}
\end{figure}

From the BP position time series alone the investigation on the system is performed at a finite resolution $r$; moreover, the dimensionality of the phase space is lost, since $X(t)\in\Re$. To cope with these limitations, the phase space is reconstructed by delay embedding technique~\cite{Kantz} in which, from the scalar time series $\{X(t_{\alpha})\}$, a new vector time series is defined as
\be
{\bf X}^{(m)}(t_{\alpha})=\{X(t_{\alpha}),X(t_{\alpha}+\tau),\ldots,X(t_{\alpha}+(m-1)\tau)\},
\ee
\noindent
being a portion of the discretized trajectory with step $\tau$ and defined in ${\Re}^m$; $m$ is known as the embedding dimension. Next, the space ${\Re}$ is partitioned using cells of length $r$, which defines a resolution for the $X$ variable. The vector ${\bf X}^{(m)}(t)$ is coded into a word of length $m$,
\be
{\bf X}^{(m)}(t_{\alpha})\rightarrow W_{m}(r,t_{\alpha})=(i(r,t_{\alpha}),\ldots,i(r,t_{\alpha}+(m-1)\tau)),
\ee
where $i(r,t_{\alpha}+j\tau)$ labels the cell in $\Re$ containing $X(t_{\alpha}+j\tau)$. Under the hypothesis of stationarity the probabilities $P(W_{m}(r))$ of the admissible words $\{W_{m}(r)\}$ are obtained from the time evolution of ${\bf X}^{(m)}(t_{\alpha})$. Then the $(r,\tau)$ entropy per unit time $h^{(m)}(r,\tau)$, a generalization of $h_{_{KS}}$ for variables measured with finite resolution~\cite{Gawa}, is defined by
\be
h^{(m)}(r,\tau)=\frac{1}{\tau}[H_{m+1}(r,\tau)-H_m (r,\tau)],
\ee
where $H_m$ is the $m$-block entropy
\be
H_m (r,\tau)=-\sum_{\{W_{m}(r)\}} P(W_{m}(r)) \ln P(W_{m}(r)),
\ee
which was computed by means of the Grassberger-Procaccia method~\cite{Grassberger}. The result for the $\epsilon$ values reported in Fig.\ \ref{fig:Pos} is displayed in Fig.\ \ref{fig:Hrtau}.

As can be readily seen, the shape of $h^{(m)}(r,\tau)$ is very similar for the weakly chaotic case $\epsilon=0.01$ and the strongly chaotic case $\epsilon=10$. This result can be readily explained by the fact that the BP, in both dynamical regimes, performs Brownian motion~\cite{RomeroBraun}. Now, a Brownian signal is a stationary Gaussian process characterized by a power spectrum $S(\omega)\propto\omega^{-2}$, which allows to obtain an explicit form for the ($r,\tau$) entropy in the limit $\tau\rightarrow0$ as $h^{(m)}(r,\tau)\sim D/r^2$, where $D$ is the diffusion coefficient~\cite{Gawa}. This asymptotic behavior is also displayed in Fig.\ \ref{fig:Hrtau} for each $\epsilon$ value considered using the numerically computed values of the diffusion coefficient in each dynamical regime~\cite{RomeroBraun}. As in Refs.~\cite{Gaspard,Dettmann1,Cencini} we considered different delay times $\tau$ to verify that the obtained result is indeed independent of the specific delay time value employed~\cite{Cencini}. From the figure it is evident that this power law scaling for $h^{(m)}(r,\tau)$ is approximately obtained already for $\tau=1$, with no noticeable differences as $\tau$ increases. This result can be explained by the fact that the natural time unit $\tau=1$, employed as the lowest delay time, is very close to the value of the fastest period of the harmonic part of the chain, $T_{\mathrm{min}}=\pi$, which is the characteristic microscopic time scale of the system; thus the employed sampling time has an adequate resolution to probe the dynamics of the BP position. From the results displayed in Fig.\ \ref{fig:Hrtau} we can conclude that the stochastic behavior of the obtained signals is indeed independent of the underlying dynamics. In fact, as will be latter explained, the mechanisms that explain the stochastic-like behavior for both $\epsilon$ values are well known for both dynamical regimes. Since our results are consistent with those of Refs.~\cite{Dettmann1,Cencini}, it is clear that no microscopic information can be obtained from the position time series of the heavy impurity.

\section{Momentum time series analysis}

\begin{figure}
\includegraphics[width=0.98\linewidth,angle=0.0]{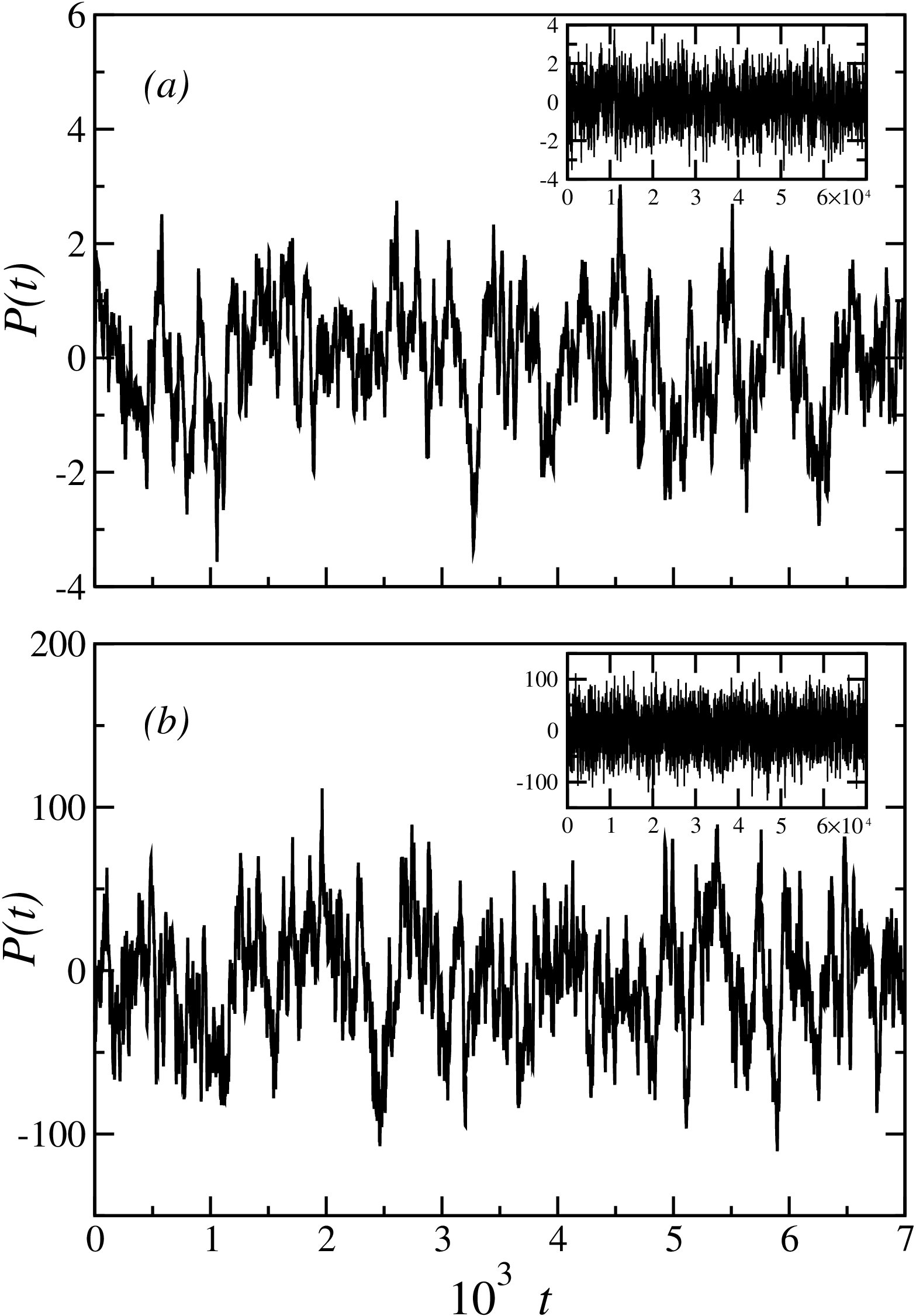}
\caption{(a) BP momentum time record for $\epsilon=0.01$. The length of the complete data set is $2\times10^5$ and the data were sampled with $\tau=1$. (b) BP momentum time record for $\epsilon=10$. Same total data set length and $\tau$ value as in (a). In both cases $M=100$ and the inset figures display the BP momentum time record over a larger time scale.}
\label{fig:Mom}
\end{figure}

Now we consider the momentum time series of the heavy impurity, which is displayed in Fig.\ \ref{fig:Mom} for $\epsilon=0.01$ and $10$. The first feature that stands out in comparison with the corresponding position time series is that the time scale in which significant variations in the signal value ocurr is much shorter than the corresponding time scale of the position variable. The second feature is that no quasirecurrences can be found by inspecting the complete time series (not shown), in sharp contrast with what happens with the complete position time series. Of course, for a long enough record quasirecurrences could be found, since $N$ is finite, as already mentioned. However, the fact that for the same data length the position variable shows quasirecurrences and the momentum time series does not clearly indicates that it may be possible to extract more dynamical information from this variable for both $\epsilon$ values considered than from the corresponding time record of the BP position variable.

\begin{figure}
\includegraphics[width=0.95\linewidth]{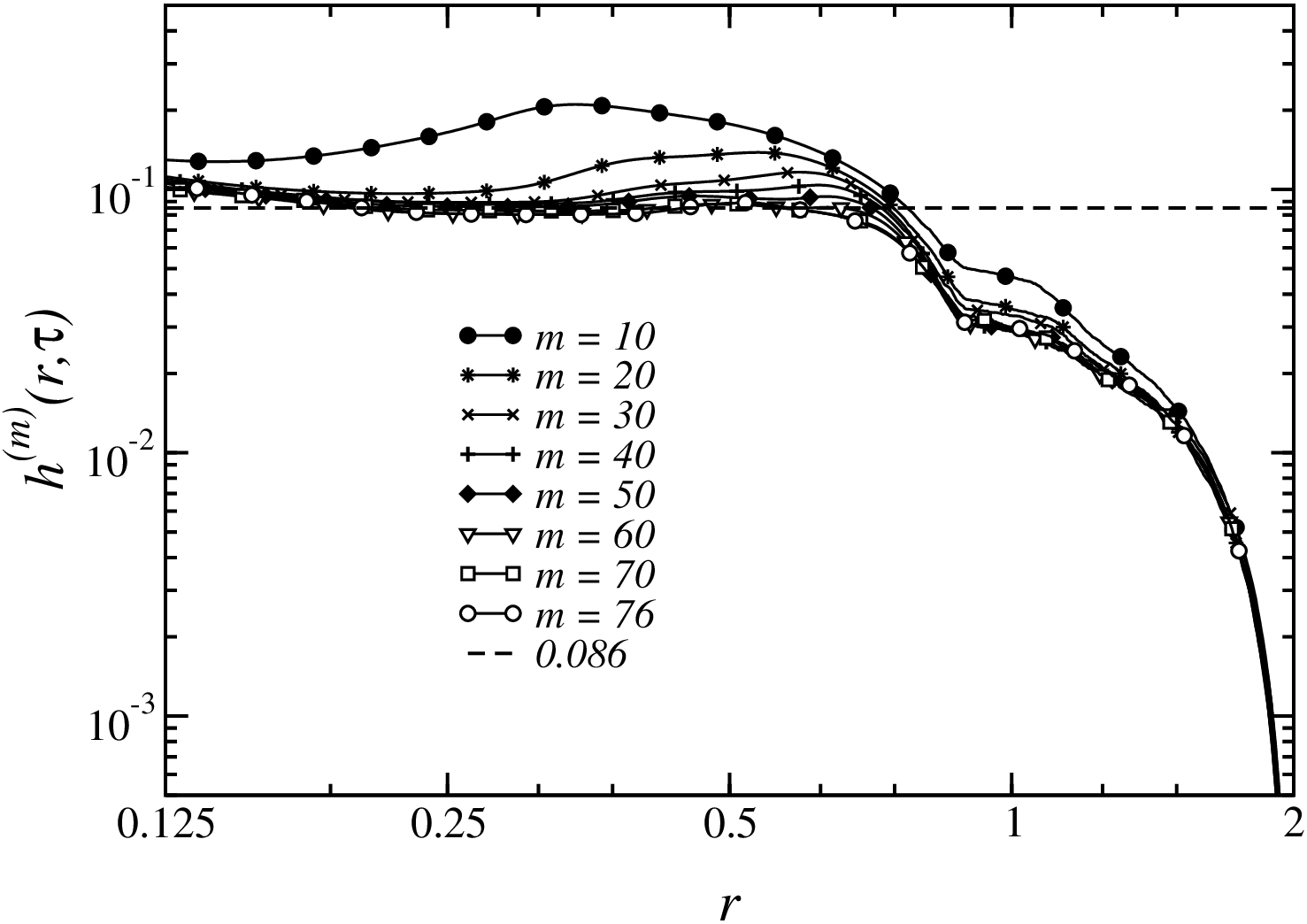}
\caption{$h^{(m)}(r,\tau)$ computed using $7\times10^4$ data points from the BP momentum time series shown in Fig.\ \ref{fig:Mom}(a). The reported results are for a delay time of $\tau=1$ and various values of the embedding dimension $m$. The horizontal dashed line is the mean value obtained from the average of the curves with $m\ge40$ for $r$ values in the range $0.19\le r\le0.59$.}
\label{fig:Hrtau_m}
\end{figure}

As in the case of the position, a reconstruction of the phase space is performed from the BP momentum time series, which yields
\be
{\bf P}^{(m)}(t_{\alpha})=\{P(t_{\alpha}),P(t_{\alpha}+\tau),\ldots,P(t_{\alpha}+(m-1)\tau)\}.
\ee
\noindent
The $(r,\tau)$ entropy per unit time for $\epsilon=0.01$, obtained from the vector time series ${\bf P}^{(m)}(t_{\alpha})$, is displayed in Fig.\ \ref{fig:Hrtau_m} for various $m$ values. The most striking feature is that the $(r,\tau)$ entropy has a well defined saturation value $h^{(m)}(r,\tau)\approx{\mathrm const.}$, which is reached asymptotically as the embedding dimension $m$ increases, for a small resolution $r$. Furthermore, as can be readily seen, there is no need to take large delay time values in order to obtain the plateau value displayed. The finite value of the $(r,\tau)$ entropy is consistent with the fact that the LLE, and thus the $h_{_{KS}}$ entropy, are well defined quantities in the thermodynamic limit, both for the FPU chain~\cite{Livi} and for the MFPU model~\cite{Romero}. It is also clear that there is no way to obtain the aforementioned plateau value using the position time series alone, as shown in Fig.\ \ref{fig:Hrtau}. The behavior of the ($r,\tau$) entropy for low $r$ values, displayed in Fig.\ \ref{fig:Hrtau_m}, is a confirmation that the momentum of the BP is a better probe of the microscopic dynamics of the oscillator chain than its position. However, the saturation value $h^{(m)}(r,\tau)\approx0.086$ obtained for $r<1$ is higher than the true microscopic value $h_{_{KS}}\approx0.12\times10^{-4}$~\cite{Romero}. 

\begin{figure}
\includegraphics[width=0.95\linewidth]{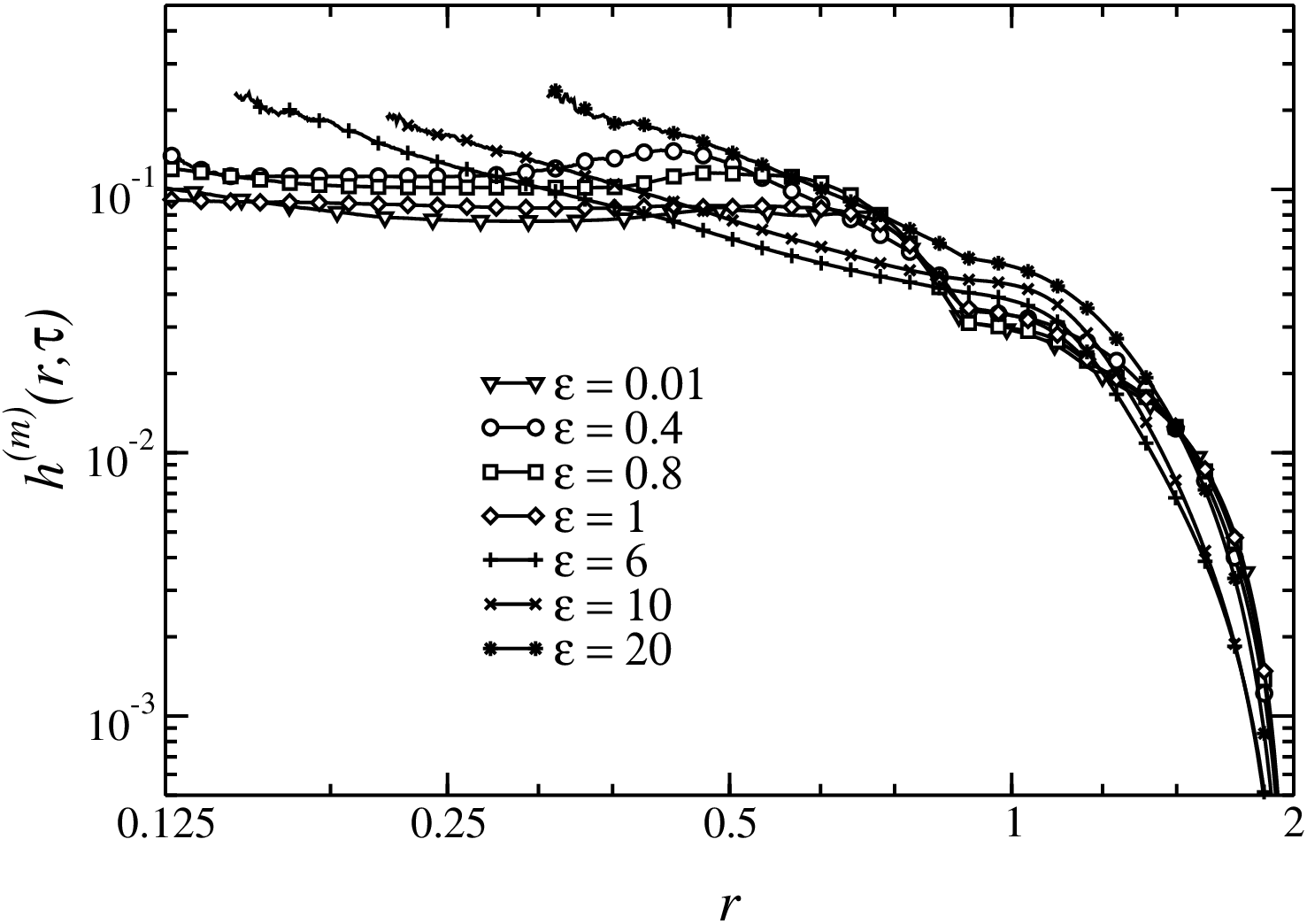}
\caption{$h^{(m)}(r,\tau)$ computed using $5\times10^4$ data points from the BP momentum time series for different energy density values. The reported results are for an embedding dimension of $m=60$, with $M=100$ in all cases.}
\label{fig:Hrtau_e}
\end{figure}

In Fig.\ \ref{fig:Hrtau_e} we present the ($r,\tau$) entropy for different values of the energy density $\epsilon$ for a single delay time $\tau=1$ and $m=60$ in all cases, since this is a representative value of the embedding dimension, as can be seen in Fig.\ \ref{fig:Hrtau_m}. We observe that the curves are separated into two different and well-defined regions, depending on the $\epsilon$ value. In all cases a plateau value can be defined when $\epsilon\le1$. On the contrary, when $\epsilon>1$, it becomes increasingly difficult to find a region in which a plateau value can be observed. So, from this figure we can establish that $\epsilon_{_T}\approx1$ is a threshold between two different regimes that are well characterized by the behavior of $h^{(m)}(r,\tau)$, as already explained. The observed difference in the behavior of the ($r,\tau$) entropy can be readily interpreted as a macroscopic manifestation of the structural changes that the phase-space undergoes when going from a low to a high $\epsilon$ value. Further details will be given in the next section.

\section{Discussion}

Our results shed some new light on the problem of detecting features of the underlying microscopic dynamics by means of macroscopic measurements. Since we have access to both the complete microscopic dynamics of the system and the results of data analysis of the BP position and momentum times series, we can make comparisons between ``exact'' microscopic information and macroscopic ``measurements'' which are not possible to make if other models are employed, as already mentioned in the Introduction. For a deterministic chaotic system one has $0<h_{_{KS}}<\infty$ ($h_{_{KS}}=0$ for regular motion), whereas for a random process $h_{_{KS}}=\infty$. Now, considering the ($r,\tau$) entropy computed from the BP position time series, there is hardly a way of discerning from Fig.\ \ref{fig:Hrtau} whether the data were originated by a an almost periodic or chaotic system, since the behavior of $h^{(m)}(r,\tau)$ is very similar in both cases. The problem can be traced to the embedding dimension $m$, since its value is well bellow the dimensionality $2(N+1)=600\,002$ of the underlying dynamical system. This is an intrinsic limitation of the nonlinear time series methods which has been previously established~\cite{Dettmann1,Cencini}. The aforementioned problem persists, but is less severe, when the BP momentum time series is employed in the computation of $h^{(m)}(r,\tau)$, since a well defined plateau value is obtained. This result can be explained by the fact that the BP momentum varies on a smaller time scale than the corresponding BP position time scale, thus allowing a more detailed sampling of the phase space. The problem becomes the overestimation of $h_{_{KS}}$ when it is approximated by the ($r,\tau$) entropy computed both from the BP position and momentum time series. The Kolmogorov-Sinai entropy is an upper bound to the $h^{(m)}(r,\tau)$ value, and thus the relation $h^{(m)}(r,\tau)\le h_{_{KS}}$ is not satisfied, as can be readily seen in Figs.\ \ref{fig:Hrtau} and \ \ref{fig:Hrtau_m}. Since the length of both the BP position and momentum time series is the same we can conjecture that this overestimation is a consequence of the small value of the embedding dimension $m$. Finally, we observe that the rate of decrease in the plateau value of the ($r,\tau$) entropy as $m$ increases becomes too small to justify any further increment in the embedding dimension, which is computationally expensive.

From Fig.\ \ref{fig:Hrtau_e} it is clear that the saturation regime of the ($r,\tau$) entropy becomes increasingly harder to define as the $\epsilon$ value is increased. Therefore two regions can be unambiguously defined by a transition from one type of behavior to another. From a macroscopic, i.e. statistical, perspective it is not at all clear what kind of mechanism could be held responsible of this change in the behavior of the ($r,\tau$) entropy as $\epsilon$ is increased. Furthermore, this transition sets in at a very precise threshold value $\epsilon_{_T}\approx1$, which in Ref.~\cite{Pettini} was called the {\sl strong stochasticity threshold} (SST). This threshold indicates a transition between two different regimes in the microscopic dynamics of the FPU chain: it is strongly chaotic and phase-space diffusion is fast when $\epsilon>\epsilon_{_T}$, whereas is is only weakly chaotic (i.e. almost periodic) and phase-space diffusion is slowed down when $\epsilon<\epsilon_{_T}$. Then, from a microscopic Hamiltonian perspective, the change of behavior detected by means of the ($r,\tau)$ entropy can be straightforwardly interpreted as a macroscopic manifestation of the aforementioned transition described by the SST. We can say that the difference in the behavior of the ($r,\tau$) entropy for low and high $\epsilon$ values means that some distinction between different dynamical regimes has indeed been observed by nonlinear time-series analysis.

Now, it is important to recall that, in the particular case of $\epsilon=0.01$, the $h_{_{KS}}$ entropy has a finite, albeit very small, value~\cite{Romero}. The apparent dynamical randomness, i.e. finite ($r,\tau$) entropy value, of the Eherenfest model has been previously explained by the structural disorder of the model~\cite{GaspardR}, which is the static analog of our random initial conditions. Thus it is quite plausible that the finite information amount carried in them is reflected in the finite ($r,\tau$) entropy value displayed in Fig.\ \ref{fig:Hrtau_m}, which corresponds to the weakly chaotic regime $\epsilon=0.01$. The increase in the ($r,\tau$) entropy of the MFPU model for high $\epsilon$ values displayed in Fig.\ \ref{fig:Hrtau_e} can be explained by the fact that it has been known for some time that the dynamics of the FPU model, for high $\epsilon$ values, mimics a random process. Thus the LLE scaling exponent, which characterizes its power-like dependence on $\epsilon$, can be obtained by means of a random matrix approximation~\cite{Pettini}. This dynamical randomness produces an increase in the ($r,\tau$) entropy relative to its plateau value for $\epsilon=0.01$. The picture that finally emerges is that, due to the small value of the embedding dimension $m$, a finite value of the ($r,\tau$) entropy will most likely be obtained when an observable of a high-dimensional system like the MFPU model is examined with nonlinear time series analysis techniques, independently of the precise nature of its microscopic dynamics, either regular or chaotic. Nevertheless, a distinction can be made between those regimes by inspection of the low $r$ behavior of the ($r,\tau$) entropy, which has a well defined plateau value for the weakly chaotic regime and a monotonically increasing value for the chaotic regime. It will be a matter of future research to explore if this type of distinction can be made in other models.

We would want to end this section with some final remarks. The central argument of this paper is that the information obtained from a given system can be very different depending on the dynamical variable chosen to be studied by time-series analysis methods. Thus, we have found evidence of the microscopic dynamics of the FPU chain by employing the BP momentum instead of the position. Now, as pointed out in Ref.~\cite{RomeroBraun}, the transition from an almost periodic to a chaotic behavior in the FPU chain has been observed in a change in the scaling behavior of the relaxation time of the MACF of the BP coupled to the oscillator chain. This is a quantity that can be readily measured in an experimental setup by means of standard techniques, such as neutron scattering~\cite{Boon}. Thus, it could be possible to explore the microscopic dynamics of more complicated systems by means of this variable, instead of using time-series analysis techniques. Within the scope of these latter techniques there remains the possibility to estimate the $(r,\tau)$ entropy by observing faster processes such as Johnson thermal noise in an electrical circuit, for example. In this particular case the analog of the BP velocity would be the electrical current. In Brownian motion the stochastic behavior of the BP velocity is produced by the interaction of the BP with the fluid particles. In Johnson noise interaction between the conducting electrons and the thermally vibrating atomic lattice of the wire give rise to a temporally varying electromotive force that is the analog of the stochastic force that appears in the Langevin equation of Brownian motion. If a suitable microscopic model of the random electromotive force could be developed, a theoretical analysis along the lines of this paper could be performed, with results readily verifiable experimentally, since the wire current is a directly measurable quantity. Of course, any other process defined by a Langevin-type equation in which a microscopic model of the corresponding fluctuating force could be provided would be adequate to explore the microscopic dynamics by means of macroscopic measurements.
 
\section{Conclusions}

In this work we have presented positive evidence, obtained from a systematic study employing time-series analysis methods, that certain qualitative information on the microscopic dynamics of a FPU chain can be detected by analyzing the momentum time-series of a heavy impurity coupled to the chain. This detection was made possible by the fact that the BP momentum varies in a much shorter time scale than the BP position. Since the BP momentum obeys a Langevin equation, our results open the possibility that a similar methodology could be applied to other processes if a certain variable that obeys a Langevin-type equation could be properly defined and a suitable microscopic dynamical model to account for the macroscopic randomness could be provided.

\begin{acknowledgments}
We are grateful to Massimo Cencini for his suggestions in the early stages of this research. We also thank the referee for his valuable suggestions and references. Financial support from CONACyT, M\'exico is also acknowledged.
\end{acknowledgments}


\end{document}